# Visual Multi-Metric Grouping of Eye-Tracking Data

Ayush Kumar
Stony Brook University, New York

Rudolf Netzel
University of Stuttgart, Germany

Michael Burch
Eindhoven Uni. of Tech., Netherlands

Daniel Weiskopf
University of Stuttgart, Germany

Klaus Mueller
Stony Brook University, New York

We present an algorithmic and visual grouping of participants and eye-tracking metrics derived from recorded eye-tracking data. Our method utilizes two well-established visualization concepts. First, parallel coordinates are used to provide an overview of the used metrics, their interactions, and similarities, which helps select suitable metrics that describe characteristics of the eye-tracking data. Furthermore, parallel coordinates plots enable an analyst to test the effects of creating a combination of a subset of metrics resulting in a newly derived eye-tracking metric. Second, a similarity matrix visualization is used to visually represent the affine combination of metrics utilizing an algorithmic grouping of subjects that leads to distinct visual groups of similar behavior. To keep the diagrams of the matrix visualization simple and understandable, we visually encode our eye-tracking data into the cells of a similarity matrix of participants. The algorithmic grouping is performed with a clustering based on the affine combination of metrics, which is also the basis for the similarity value computation of the similarity matrix. To illustrate the usefulness of our visualization, we applied it to an eye-tracking data set involving the reading behavior of metro maps of up to 40 participants. Finally, we discuss limitations and scalability issues of the approach focusing on visual and perceptual issues.

## Introduction

Understanding participants' behavioral patterns and learning about commonalities and diversities in these is a challenging task. These types of analyses were done in past for many fields, such as usability research using machine learning, data mining (Paliouras, Papatheodorou, Karkaletsis, & Spyropoulos, 2002), grouping of users for music recommendation (Chen & Chen, 2001) based on complex amino acid sequences (Vernone, Berchialla, & Pescarmona, 2013). Such application exists also in the field of eye tracking, which is the focus of this paper. In this context, Netzel et al. (2017) conducted a user performance analysis and grouped participants based on a fixation label sequence of scanpaths and the bimodality coefficient. Kurzhals et al. (2014) grouped users on the basis of similarity functions such as Levenshtein distance applied to scanpaths, a function based on attention distribution, and one based on AOI transitions (Li, Çöltekin, & Kraak, 2010). Anderson et al. (2015) also conducted scanpath comparison in an experiment to reveal the similarities within and between the scanpaths of individuals looking at natural scenes.

West et al. (2006) used fixation sequences to highlight groups, or clusters of sequences that are similar. Traditional approaches for comparing eye-movement behavior of several study participants typically focus on the scanpaths by aligning them as good as possible to detect commonalities (Raschke, Chen, & Ertl, 2012); (Burch, Kumar, Mueller, & Weiskopf, 2016). All of these applications have shown that it is important to group users on the basis of their common behavior. However, the above-mentioned grouping approaches are only based on few standard descriptors of eye-tracking data such as fixation duration, scanpaths, saccades, AOIs, etc. From a machine learning perspective, the result of the grouping or clustering highly depends on the used feature vectors with attributes that describe characteristics of data instances and their relation. In the context of eye tracking, these attributes are metric values derived from recorded gaze data. The main challenge lies now in the selection of appropriate metrics, which is not easy to achieve, since the characteristics that the metrics should describe highly depend on the used stimuli and task. Hence, there is no fixed set of metrics that can be used to achieve a grouping of participants. Therefore, it is necessary to investigate and select the space of metrics for each new experiment.

Such evaluations and investigations are nowadays possible, since tracking people's eye can be done reliably and accurately, assuming an advanced eye-tracking system as described by Kurzhals et al. (2015), this can generate an enormous amount of data, eventually leading to big data (Blascheck, Burch, Raschke, & Weiskopf, 2015; Burch, Chuang, Fisher, Schmidt, & Weiskopf, 2017). Eye-tracking metrics (Holmqvist et al., 2011) have been explored extensively to derive meaning and statistical values from the recorded data. All of these more or less focus on different aspects of the data, as to describe properties from different perspectives. Different metrics are used to interpret behavior and derive possible cognitive states. Looking into a combination of metrics is also possible to improve the quality of results and the interpretation thereof. Consequently, the number of used metrics can become rather large and the handling of such multivariate data becomes challenging in terms of analysis and visualization.

In this paper, we provide an approach with which correlations among the metrics can be explored, or participant groups of similar behavior can be identified. To reach this goal, we first apply well-known concepts for multivariate data visualization based on parallel coordinates (Inselberg, 1985; Heinrich & Weiskopf, 2013). The eye-tracking metrics build the axes of the parallel coordinates plots, whereas the polylines are given by the individual eye-tracked people and their respective metric values under observation. By such a plot, we can identify correlations among pairs of metrics, even if the number of eye-tracked people gets rather large. Moreover, parallel coor-dinates can serve as a selection tool to further decide which metrics are of particular interest for data analysis.

Starting from such an overview of multivariate eye-tracking metrics data, we further support interaction techniques with which the analyst can have a deeper look into metric correlations. A multi-metric approach with weighted scores of various metrics of eye-tracking data, can be used for an overall grouping. Pajer et al. (2017) suggested an interactive solution to find out appropriate weighted values in multi-criteria decision-making scenario, which we have used in our work to find a suitable weight for the combination of multiple metrics to generate a weighted score to be employed for clustering. Thus, a multi-metrics-based approach, as in this work, is scalable and can provide different views on the eye-tracking data by selecting a list of crucial metrics out of many. Instead of analyzing the eye-tracking data based on an individual metric, we furthermore support a combined view on a set of metrics. Such multi-metric clustering is useful to indicate participant groups that behave similarly with respect to observed eye-tracking metrics. Unfortunately, parallel coordinates alone allow just the pairwise direct comparison of axes (i.e., eye-tracking metrics). Therefore, an additional interactive visualization is required that shows as many metric values as selected for each participant in combination.

To mitigate the problems that parallel coordinates have, we provide a second view based on matrix representations. This allows us to see the outcomes of a multi-metric grouping based on clustering and, hence, provides insights into commonalities of eye-tracking patterns of a group of people. The metrics-based clustering of participants into groups works fast and supports interaction. Another benefit of a matrix-based visualization is reduced visual clutter (Rosenholtz et al. 2005), since the data is mapped to rectangular, possibly pixel-based, graphical primitives. This is different from parallel coordinates, which result in line-based diagrams. In case of a matrix-based visualization, grouping of participants can directly be integrated from a dendrogram that shows the hierarchical organization of the gaze data based on one or more user-selected eye-tracking metrics.

This article is an extension of a formerly published research (Kumar et al. 2016) focusing on the matrix visualization and a clustering approach, while only a few metrics were involved in the clustering and visualization process. Inspired by the comprehensive presentation by Holmqvist et al. (2011), we extended the previous list of eye-tracking metrics and support a preselection of those by means of a parallel coordinates plot in which the axes can be interactively chosen by a data analyst. In this paper, we add the following contributions:

- **Parallel coordinates:** To get an overview, we have formulated eye-tracking data as multivariate eye-tracking metrics. To show correlations between the metrics, we provide parallel coordinates plots enhanced by interaction techniques like brushing and linking, value range selections, weighted metric combination, and participant group color coding.

- **Matrix visualization:** A complementary view on the multivariate eye-tracking metrics is provided by a matrix visualization that encodes the metrics in the matrix cells as a pairwise comparison between the participants. The similarity values for the pair of participants are then ordered based on correlation coefficients and a Hilbert space-filling curve to preserve locality.

- **Interaction techniques:** Further interactions are integrated that connect both visual representations, i.e., the parallel coordinates reflecting metrics correlations, but also the matrix visualization, depicting participants groups with similar behaviors based on multiple eye-tracking metrics.

**More advanced data analysis:** In particular, we extended the tool by adding further techniques like hiding axes (metrics) of less importance, merging those axes into one, and deleting data points of less interest.

We illustrate the usefulness of our technique by applying it to eye-tracking data from study that investigated a route-finding task in public transportation systems (Netzel et al., 2017).

Finally, we discuss scalability issues based on visual and perceptual problems coming with our multi-metric grouping. The framework used in this paper including all codes can be downloaded from https://github.com/hawkeye154/JEMR17.

## Methods

In this section, we are going to describe the used techniques as well as our visual analytics workflow model that is set up in three stages (see Figure 1): data preprocessing, metric analysis, and participant group analysis. All generated metrics, after preprocessing the raw gaze data in a first phase, are plotted as parallel coordinates. An analyst supported by various interactive techniques provided by the framework, can then explore the parallel coordinates for the ultimate goal of grouping participants. However, the visual clustering done for grouping participants in parallel coordinates is not very clear. So, the feedback provided by parallel coordinates in the metric analysis phase is used in the participant group analysis phase and vice versa. The participant group analysis phase is supported by a similarity matrix visualization based on algorithmic clustering, where metric information is used for the generation of good clusters as proposed by Kumar et al. (2016). For a matrix-based visualization, we calculate the similarity values for each pair of participants. These values are then used for clustering as well as visualization.

### Data Model and Preprocessing

To illustrate and validate our visualization tool, we have chosen an eye-tracking data with 40 participants each looking at 48 stimuli of metro maps for study (Netzel et al., 2017). This data set is a typical representation of eye-tracking data consisting of fixations and saccades along with their coordinates, and time stamps after a fixation filter was applied on the raw gaze data. It consists of a sequence of fixations (scanpath) for each participant $P_i$ and each stimulus (metro map) $S_j$. Based on this data, we derived eye-tracking metrics. We have used all stimuli as well as all participants data and considered participants as a variable, since we are interested in grouping participants with common behavior on the basis of their eye-tracking data. However, if an analyst is interested in grouping stimuli based on the performance of participants, then the stimuli are considered as a variable.

We have used metrics from a large range of measures discussed in the book of Holmqvist et al. (2011) and selected typical metrics from three categories of measures, i.e., eye-movement measures based on saccades, numerosity measures by counting events, and statistical (position) measures of the spatial distribution as shown in Table 1. While metrics from the first two categories are often used for evaluation, simple statistical measures could be helpful to get an aggregated view of the positions of fixations. Skewness indicates whether fixations are located on the left or right (top or bottom) side of the stimulus. Standard deviation indicates the general spread of the fixations. Finally, kurtosis tells us whether there is a tendency for multiple clusters of fixations or rather a singular occurrence.

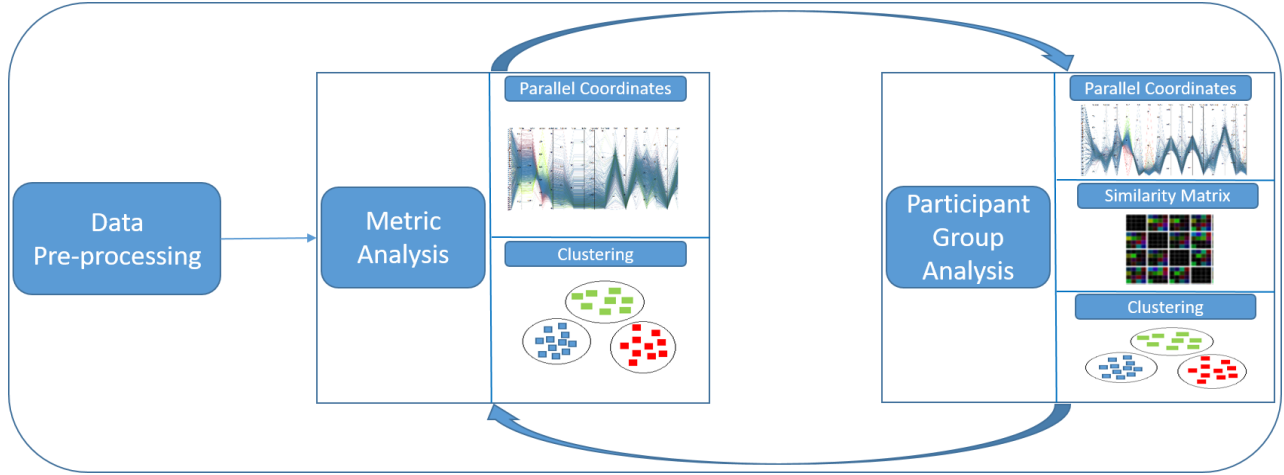

*Figure 1.* Work-flow overview: Preprocessed data are fed into parallel coordinates plot for visual clustering and providing grouping information that can be used in the similarity matrix visualization. Gained insights serve to steer the analysis based on parallel coordinates and algorithmic clustering.

For ease of representation, we call the metrics $M_k$, where $1 \leq k \leq n$. For each metric, we calculate the similarity values $sv_{M_k}$ for each pair of participants $P_{i,l}$. There are $|P| = p$ participants in total. Hence all similarity matrices will be of size $p \: x \: p$. $\frac{p!}{2!(p-2)!} = \binom{p}{2}$ different similarity values, one for each pair of participants, and a total of $\binom{p}{2}n$ for all $n$ metrics. We calculate similarity values $sv_{i,l,M_k}$ between two participants $P_i$ and $P_l$ based on metric $M_k$ using the Euclidean distance (Equation 1) to find similarities between the reading behavior of participants based on their eye-tracking data.

$$sv_{i,l,M_k} = \left\| P_{i,M_k} - P_{l,M_k} \right\|_2 \qquad (1)$$

In our example, $sv_{i,l,M_k}$ are scalar values, which resulted into $n$ similarity matrices of size $p \: x \: p$. To make it worth for visual inspection, we combined all the similarity values of $n$ metrics into a single matrix as stacked rectangular sub-grid as proposed by Kumar et al. (2016).

Details will be discussed in the section on *Similarity Matrix Visualization*.

To demonstrate the usefulness and application of our visualization framework for exploration of eye-tracking data, we have computed 16 metrics according to Table 1.

## Visualization and Visual Analytics

For the visual exploration of eye-tracking data, we use clustering and two different visualization techniques in combination. Due to the large number of metrics, we chose to use parallel coordinates, which are popular for the visualization of high-dimensional data. They are also employed to visually cluster and explore data, which gives an analyst an idea of the metrics that could be used for a grouping of participants. Clustering is then applied to the multi-dimensional stacked matrix as discussed in the section on *Similarity Matrix Visualization*.

***Clustering.*** Clustering groups of objects or data points based on similarity values helps find structures within data or relations between objects. We have used agglomerative hierarchical clustering of participants based on the metrics used in this study. It is one of the main stream clustering methods with a complexity of $O\left(N^2 \log N\right)$. The popularity of this clustering is due to the fact that it does not need any predefined parameters, which makes it easier to be used for all sorts of real-world data (Bouguettaya, et al., 2015).

The main challenge faced during clustering is choosing the right combination of similarity values to be used for clustering. Even after finding correlation between metrics, it is difficult to identify eye-tracking data metrics that will be beneficial for grouping. Performing clustering for all the combination of metrics for grouping participants could be very tedious.

To solve this problem of finding suitable metrics for clustering, we use parallel coordinates to drill down into the metrics and their properties by various interaction techniques described in the following sections, as parallel coordinate plots can be used for visual clustering, i.e, to find groups based on visual features. We have tried various combinations of metrics through one of the interactive technique of our tool providing varying weights to each of them. It gives us an approximate idea about which metrics we can use in combination for calculating the similarity value matrix. This similarity matrix will then be used to group participants by clustering.

***Parallel Coordinates.*** Understanding complex data has always been a challenging problem, and there are various visualization techniques to understand the data. There are several standard visualization techniques to visualize data such as histograms (1-dimensional) or scatterplots (2-dimensional). The problem becomes even pronounced when we have multi-dimensional data (Le-Blanc, Ward, & Wittels, 1990; Fua, Ward, & Rundensteiner, 1999). In order to visualize data with such techniques, we need to produce several plots, which is not feasible with an increasing number of dimensions.

To show a lot of data dimensions, we chose parallel coordinates for visualizing eye-tracking metrics for the second part of our visual analytics workflow. Parallel coordinates made their first-time appearance in 1885 (d'Ocagne, 1885). However, they became more popular as a tool for multi-dimensional data exploration after the work of Inselberg (1985) and Wegman (1990). Parallel coordinates are based on the concept of point-line duality, where data points in Cartesian space are plotted in the form of lines in parallel coordinates. In parallel coordinates, two or more axes are plotted parallel to each other, where each axis serves as one dimension of the multi-dimensional data. Data points are then plotted on the parallel axes, which results into intersection of polylines at respective data values, as shown in Figure 2(a). Multiple data points can be plotted similarly, as shown in Figure 2(b).

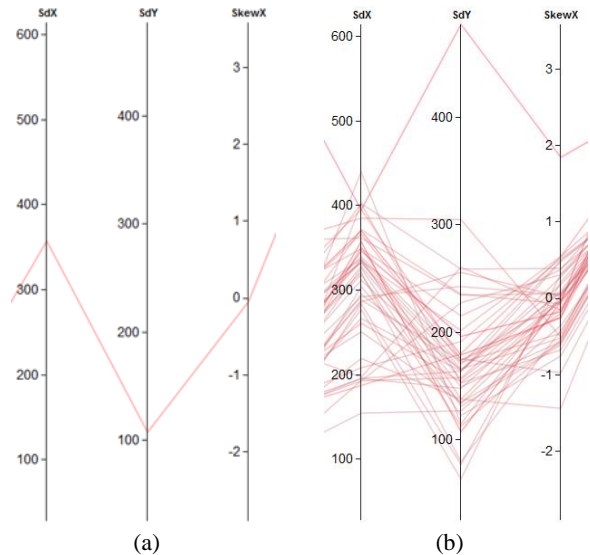

*Figure 2.* Parallel coordinates for multiple data dimension: (a) polyline plot of one data point and (b) polylines of the multiple data points.

Parallel coordinates plots serve the purpose of visual exploration, as finding the relation among the eye-tracking metrics in our case. Of all the three possible correlations (positive, negative, and none), negative correlation has the most striking pattern. The intersection of data points of higher values on one axis are mapped to lower values on the neighboring axis and vice versa, resulting into an accumulation point that is easily spotted. A sample case for perfect negative correlation is shown in Figure 3(a). Positive correlation shows properties opposite of negative correlation, thus resulting in straight lines, as depicted in Figure 3(b).

There are several interaction techniques that support visual exploration. Our parallel coordinates framework, as shown in Figure 4, provides a collection of such interaction techniques. Brushing, being one of the basic action techniques for parallel coordinates plots, allows the user to select data points on the axis. The selection is typically done in an interactive fashion; often, the user marks interesting areas in the plot by mouse interaction as shown in annotation (a) of Figure 4. There are generally three types of brushes, such as axis-aligned brush, polygon-shaped brush, and angular brush (Hauser, Ledermann, & Doleisch, 2002). We use axis-aligned brushing, which enables us to select points in a vertical fashion on axes at a time.

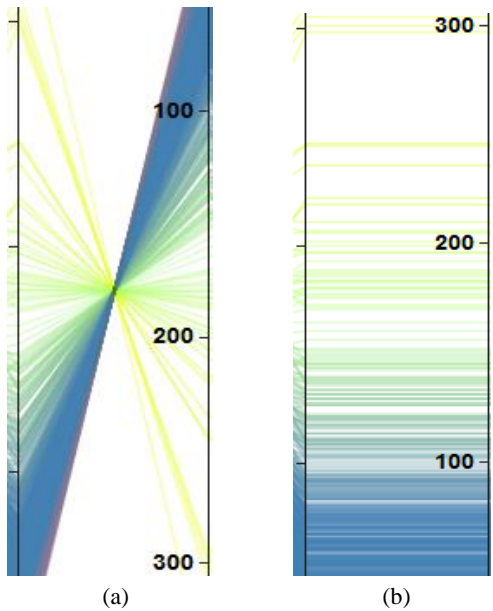

*Figure 3*. Visual patterns in parallel coordinates with (a) negative correlation and (b) positive correlation.

Selected data points are then used for further exploration. We can color each axis individually based on the data points plotted on each axis, as for an example case *SkewY* is color-coded in annotation (g) of Figure 4. The user can apply coloring according to the metric of their choice by simply clicking on that axis. To find correlations between metrics, we need to order the axes representing metrics in a meaningful manner. We can reorder an axis by simple dragging and dropping it at the desired position. Sometimes correlation can also be seen by inverting the orientation of an axis, which is supported by double clicking on an axis.

There are several problems with the polylines used in the traditional parallel coordinates plots. They result in loss of visual continuation if two or more data points have similar values, which makes it difficult to follow the line throughout the plot. To solve this problem, we optionally replace the polylines with smooth curves controlled by a slider (see annotation (c) of Figure 4) as done by Graham et al. (2003). Smoothing the polylines is beneficial for bundling as it allows analyst to take advantage of the entire plot area because it separates distinct components of data into distinct regions by assigning a centroid as shown in Figure 5(b). Curve bundling is also highlighted in our framework, which supports the formation of curve bundles for clustered data plotted on parallel coordinates, as shown in Figure 5. The slider used in the framework (see annotation (d) of Figure 4), helps the user tune the tightness of the bundled curve, which enables them to have a different view of the same data (Heinrich et al., 2011).

There could be a case where two or more metrics used for analysis are of similar nature. Our framework allows the analyst to remove axes that are not of interest by selecting check-boxes as shown in annotation (b) of Figure 4. There could be a scenario where the analyst wants to merge two or more axes, assigning desirable weight to each one of them for the metric analysis in order to group the data based on the new combined metric. To support this feature, we have a section in our framework (see. annotations (e) in Figure 4) that enables user to select the metrics they want to combine. The weight can be adjusted with a slider below the checkbox.

*Similarity Matrix Visualization.* The third part of our visual analytics process deals with the participant group analysis and uses a matrix-based approach for visualization. It enables users to distinguish groups of participants based on various metrics of eye-tracking data. A matrix-based representation is one of the most basic ways of visualizing two-dimensional data. However, it is difficult to visualize multi-dimensional data using a matrix. Therefore, we adopt the concept of dimensional stacking: embedding dimensions within other dimensions (LeBlanc et al., 1990). Each matrix cell is divided into multiple sub-grids depending on the number of metrics to be stacked as used by Kumar et al. (2016). In this paper, each grid is divided into 16 sub-grids since 16 metrics are supposed to be embedded into single grid.

Each sub-grid is used to stack the similarity values $sv_{i,l,M_k}$ calculated in Equation 1 for each metric. However, it is not easy to arrange the data with $n$ metrics into a $\lceil \sqrt{n} \rceil$ x $\lceil \sqrt{n} \rceil$ grid and to make it visually appealing, so that it becomes easier to visually group participants on the generated clustered matrix representation. We solve this issue by computing correlation coefficients between all pairs of $n$ metrics, which results into $\frac{n!}{2!(n-2)!} = \binom{n}{2}$ values. We then clustered all $\binom{n}{2}$ combinations using hierarchical clustering and plot them in the form of a dendrogram as shown in Figure 6(a). This clustering is now, used to arrange similarity metrics next to each other. However, the problem of ordering these metrics in a sub-grids is still challenging. So, we followed the Hilbert space-filling curve to fill the sub-grids, as shown in Figure 6(b). This space-filling representation keeps nearby values close to each other, which helps preserves locality (Moon et al., 2001). For our study, we chose 16 different metrics preprocessed from the raw gaze data in the first phase. Therefore, we chose 16 different colors in the CIELAB color space that represent all 16 similarity values $sv_{i,l,M_k}$ in each sub-grid as base colors, as shown in Figure 6(c). The selected 16 colors are saturated colors from the CIELAB color circle with uniformly distributed hue values, with a hue difference of 20 between neighboring hues. We chose hue to encode metric classes because hue is effective in perceptual grouping of categorical data (Ware, 2012). The CIELAB color space is well suited since colorimetric distances between any two colors in this space corresponds to perceived color differences due to its three-dimensional in nature where each a*, b* and L* respectively serves as an individual dimension. While a* and b* being chromatic channels of the CIELAB color space is fixed according to the class of metric, lightness L* varies to encode the similarity values for the respective metric. We assigned these colors to metrics in sub-grids in the order of the Hilbert space-filling curve, which resulted in Figure 6(d). An example of eight participants with 16 metrics is shown in Figure 6(e).

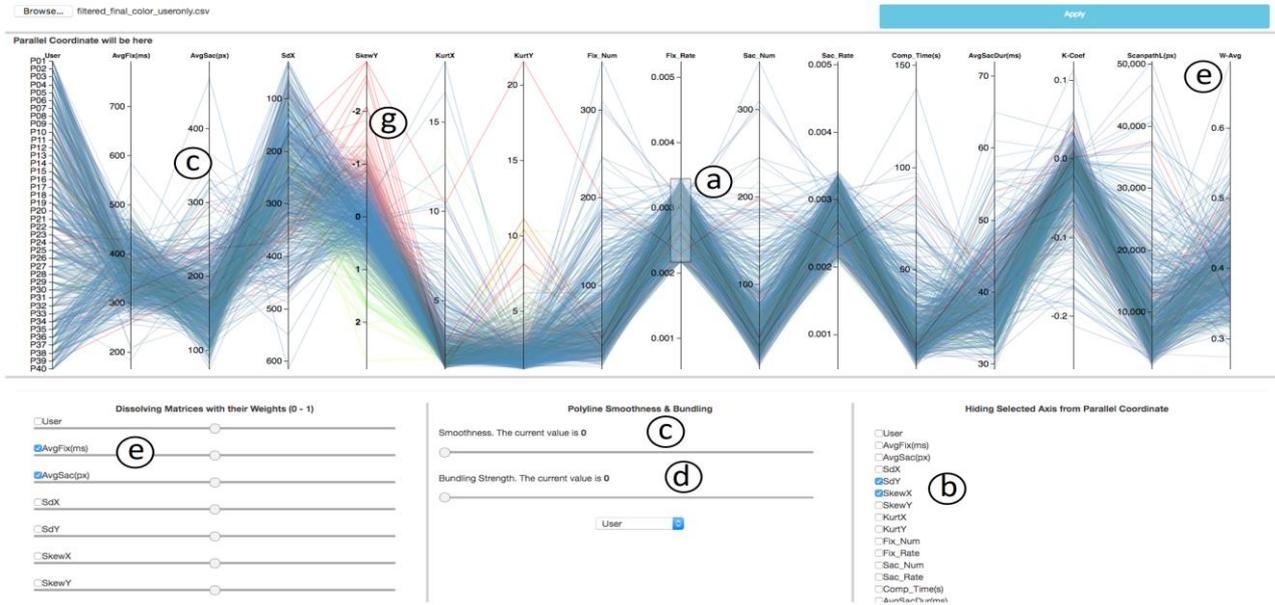

*Figure 4.* Overview of the parallel coordinates tool. Axis-aligned brushing is shown in (a). Axes can be deleted by (b), polylines that pass through an axis can be smoothed if needed (c), curves can be bundled around a centroid using this option (d), two or more metrics can be added with weights (e), color coding to distinguish the cluster based on data value in a specific metric can be applied as shown in (f).

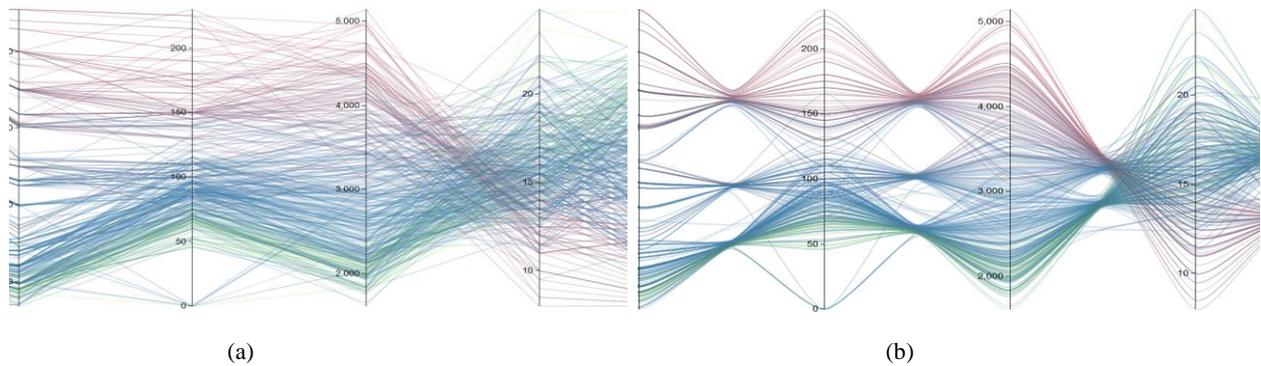

|     (a)     |     (b)     |

*Figure 5.* Parallel coordinates plot of clustered data shown in (a) and data representation with bundled curve separating distinct component shown in (b).

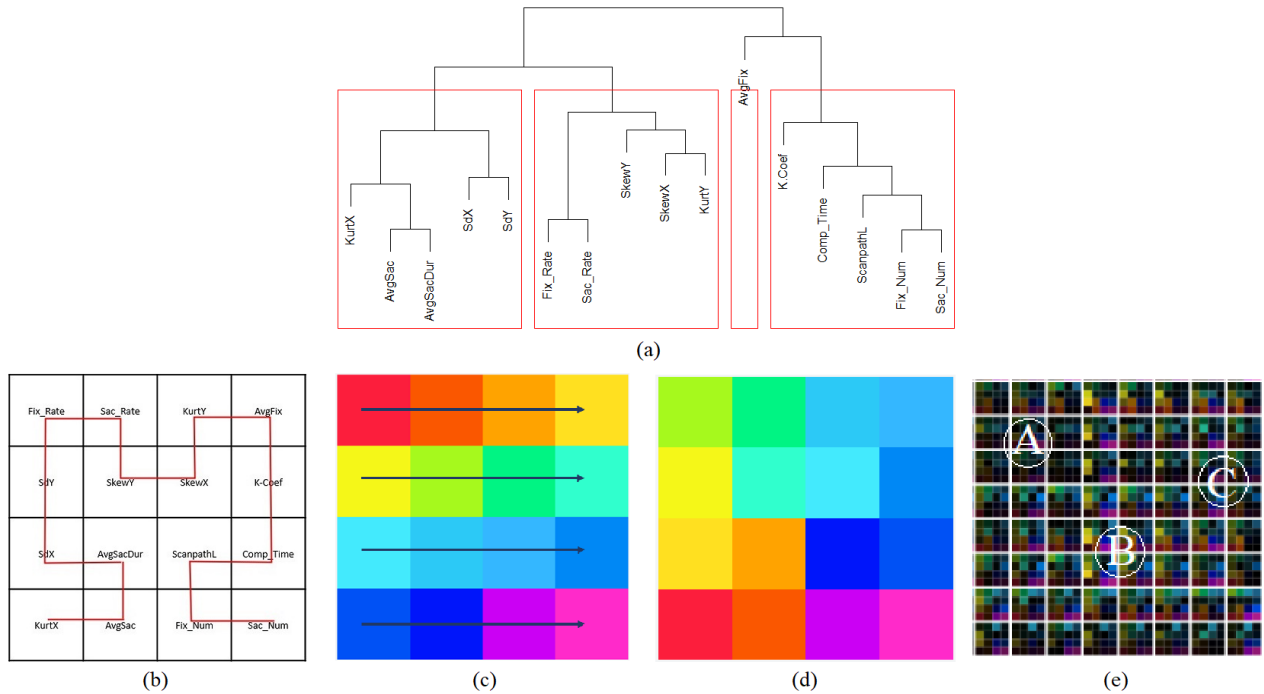

*Figure 6.* Overview of the procedure followed in the generation of Dimensionally Stacked Similarity Matrix (DSSM). The dendrogram shown in Figure (a) reveals the proximity between metrics from eye-tracking data. Figure (b) shows the order in which metrics are stacked in the similarity matrix following a second-order Hilbert curve. Figure (c) shows the 16 colors with varying hue to represent each similarity value in the sub-grid. Figure (d) shows that choosing the Hilbert curve preserves locality in colors. Figure (e) displays an example of multi-dimensional stacked metrics with dimension 8 x 8 for 8 participants.

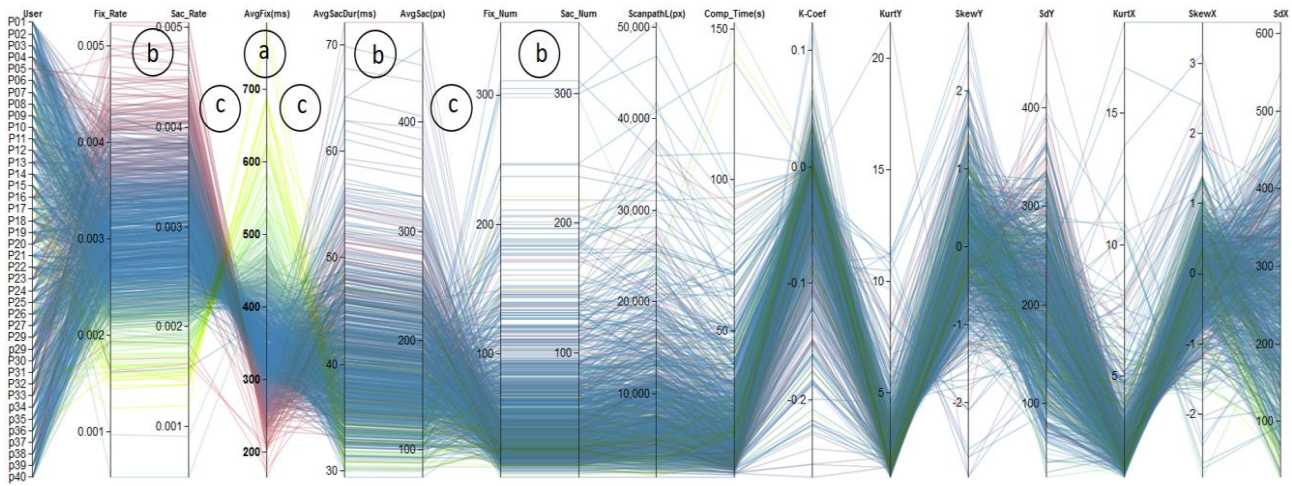

*Figure 7.* Overview of all variables in a parallel coordinates diagram. Color indicates different clusters obtained from clustering based on the fourth axis (*AvgFix*, (a)). The visualization also reveals positive correlation between attributes (b) and negative correlations (c).

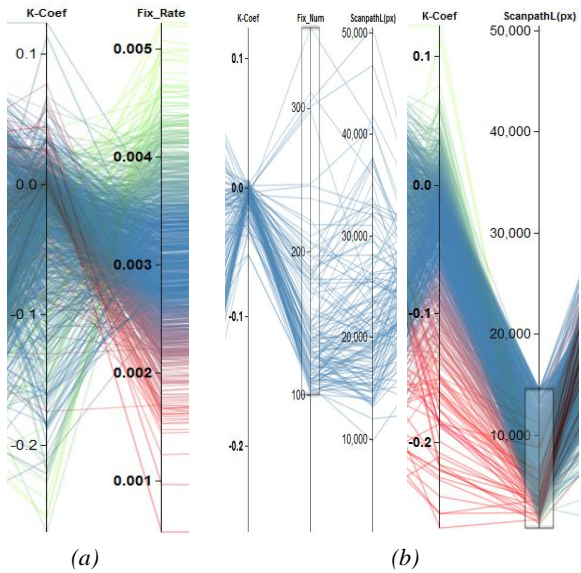

*Figure 8.* Examples of finding effects based on the clustering of a single attribute. (a) shows the negative correlation between *K Coef* and *FixRate*. (b) depicts two different groups of behavior regarding the *K Coef* based on *scanpath length*. Scanpaths longer than about 15,000 px ((b), left) lead to *K Coef* values centered around zero. Shorter scanpaths ((b), right) result in a broader negative distribution.

## Results

With the data obtained from the previously described eye-tracking study, we first derive metrics that can be used to interpret user behavior. This corresponds to the *Metric Analysis* of our established work flow in Figure 1. Next, we utilize our framework to visualize and analyze groups of participants exhibiting similar characteristics, which is related to the *Participant Group Analysis* block in Figure 1. The analysis is based on previously derived metrics.

### Utilizing our Framework

The first step in most evaluations is to generate an overview of the data that should be analyzed. Here, the data corresponds to metrics derived from recorded eye-gaze data, i.e., we first focus on the *Metric Analysis* part of the visual analytic workflow (see Figure 1). This can be shown in Figure 7, as an example of metrics data visualization using parallel coordinates. By rearranging the axes (metrics), it is possible to swiftly find those, that exhibit similar or identical behavior indicated through

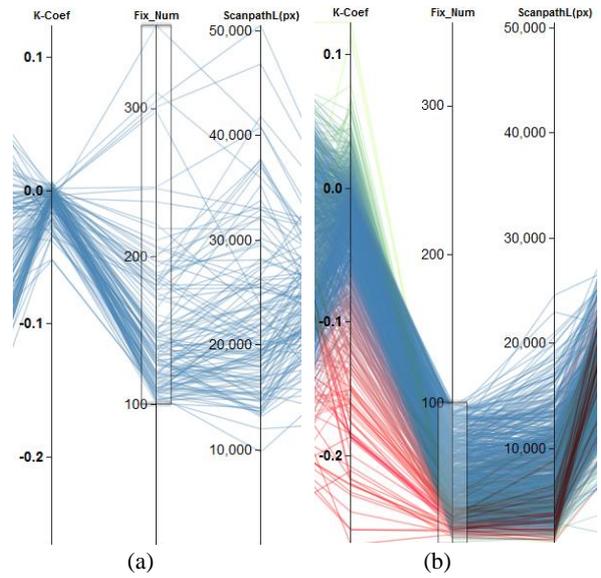

*Figure 9.* Separation of participants into two different groups of behavior regarding the *K Coef* based on the number of fixations. (a) shows all subjects with more than 100 fixations. They lead to *K Coef* values centered around zero. Fewer fixations result in a broader negative distribution, which is shown in (b). This achieves about the same separation as utilizing the *scanpath length* with a threshold of about 15.000 px.

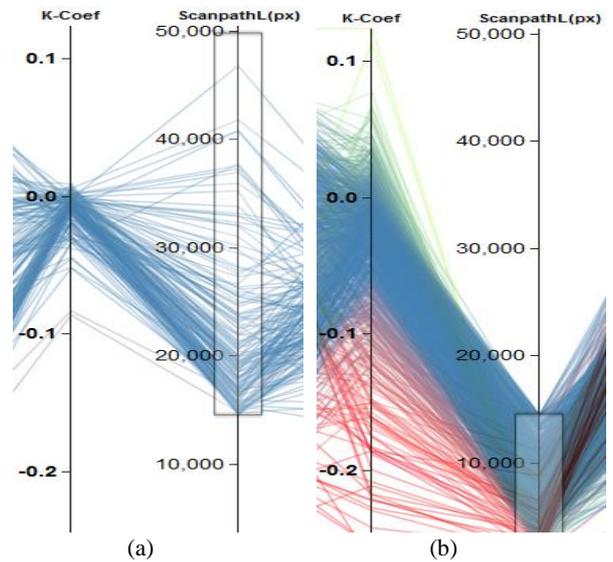

*Figure 10.* Combining completion time and scanpath length to generate a new 1D metric (*W-Avg*). (a) shows that values above a threshold of 0.3 of *W-Avg* result in one group of participants that exhibit ambiguous gaze behavior, whereas (b) depicts the second group having a more ambient gaze behavior, according to the *K* coefficient.

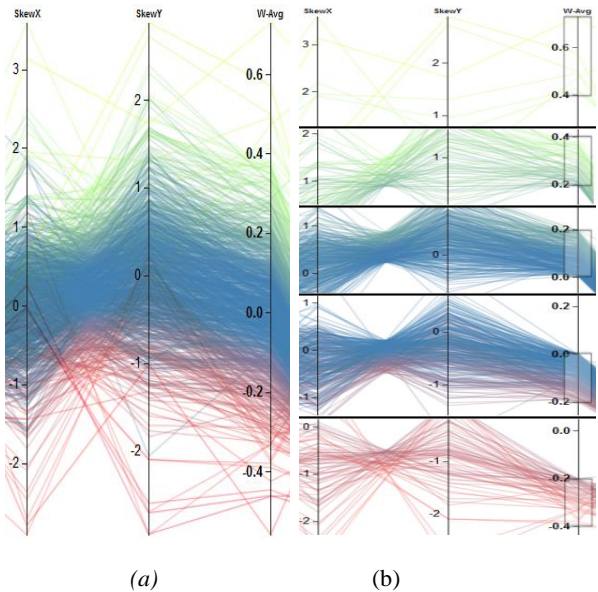

*(a)*        *(b)*

*Figure 11.* The color-encoded results are shown in (a). Clustering was performed on the weighted average of the first two columns. Overall, a negative correlation can be recognized. Selecting different values ranges of the weighted average leads to the selection of subgroups of similar skewness values. The weights were set to 0.5 for the two metrics. All subgroups exhibit also a negative correlation. This is shown in (b).

parallel line segments between neighboring axes. This is the case for *FixRate/ SacRat*, *AvgSacDur/ AvgSac*, and also *FixNum/ SacNum* (annotations (b) in Figure 7), enabling an analyst to discard metrics that convey the same kind of information. Furthermore, we are able to find metrics that exhibit an inverse behavior, which is the case for, e.g., *SacRat/ AvgFix*, *FixRate/ AvgFix*, *AvgFix/ AvgSacDur*, and *AvgSac/ FixNum*. An increase of the first attribute of each tuple leads to a decrease of the second attribute. These findings reflect the well-known relationships between the previously described metrics. Another way to find similarities between attributes is possible by selecting a specific axis that will cluster the data based on the selected metric and a color will be assigned to each data instance. Considering the color of the lines as well as the order of the color at other axes, we can see similar behavior. This is shown in Figure 7, where clustering was performed based on *AvgFix* (see annotation (a)), resulting in the same color gradient on the axes *FixRate* and *SacRate*.

Using clustering based on a single metric, we can find an inverse relationship between $\mathcal{K}$ *Coef* and *FixRate* (see

Figure 8(a)). This indicates that the ambient/focal attention changes inversely to the fixation rate. Furthermore, ambient/focal attention varies with the *scanpath length*. As seen in Figure 8 for a length of more than about 15,000 px, the distribution of the $\mathcal{K}$ coefficients is centered around zero (ambiguous cases), whereas for shorter scanpaths the distribution is more widespread and tends toward negative values (ambient attention). The $\mathcal{K}$ coefficient is also related to the number of fixations in a similar fashion. Here, we can observe the same kind of distributions for more than 100 fixations (ambiguous cases) and fewer fixations (ambient attention). If we consider the scanpath length, we can see that the separation into two groups having more and less than 100 fixations yields roughly the same separation as splitting the subjects according to a scanpath length of more or less than 15,000 px. This is shown in Figure 9.

Another interesting question is how the $\mathcal{K}$ coefficient is related to the *completion time* and *scanpath length*, since there could be, e.g., four subgroups of participants using a categorization of fast/slow for the *completion time* and short/long for the *scanpath length*. Previously, we had already a closer look at the *scanpath length* and the $\mathcal{K}$ coefficient. Therefore, we now want to focus on the influence of the *completion time,* assigning a weight of 0.7, while we assign a weight of 0.3 to the *scanpath length*. The results are shown in Figure 10. Based on this affine combination of metrics, we achieve a similar result compared to utilizing the *scanpath length* alone. One group of participants exhibits ambiguous eye movements, whereas the other groups show more ambient eye movements.

Besides the so-far-used eye-tracking metrics that characterize user behavior, we are also interested in identifying spatial behavior. Therefore, we have a closer look at metrics that give us an impression about where the participants were looking at. The skewness of the distribution of positions, projected onto the x- and y-axis, can be used for getting a rough impression. Figure 11 shows the combination of *SkewX* and *SkewY* with an equal weight of 0.5 for each metric. Overall, we can see a negative correlation between the metrics, which is also the case, if we have a closer look at different intervals of values of the combined metric (see Figure 11(b)).

To further investigate the spatial distribution, we can include kurtosis into the analysis. Based on the skewness, we have derived rough areas where participants were looking at. The kurtosis of the distributions describes now whether there are frequent small deviations of mean values or frequent larger dispersed data. The first case would indicate the existence of focus attention areas, whereas the later a wider spread of attention on possibly multiple locations. Since we are more interested in the effects of kurtosis, this metrics got a higher weight (0.7). Skewness got lower weights (0.3). Based on the new combined metric, we could roughly separate participants into four groups. Each group shows distinct visual patterns. The first one (Figure 12(a)) contains only a minority of the data that shows large changes between neighboring axes. The second group (Figure 12(b)) exhibits a drop-off of changes from left to right. The third (Figure 12(c)) and fourth (Figure 12(d)) group show similar values for the kurtosis in both spatial directions and also for the skewness in x-direction, but differ visually in the skewness in y-direction. With this kind of analysis, using parallel coordinates, we were able to investigate characteristics of each scanpath of each participant during the *Metric Analysis* step of our workflow (see Figure 1) and to form clusters based on color encoding or similar visible line structures.

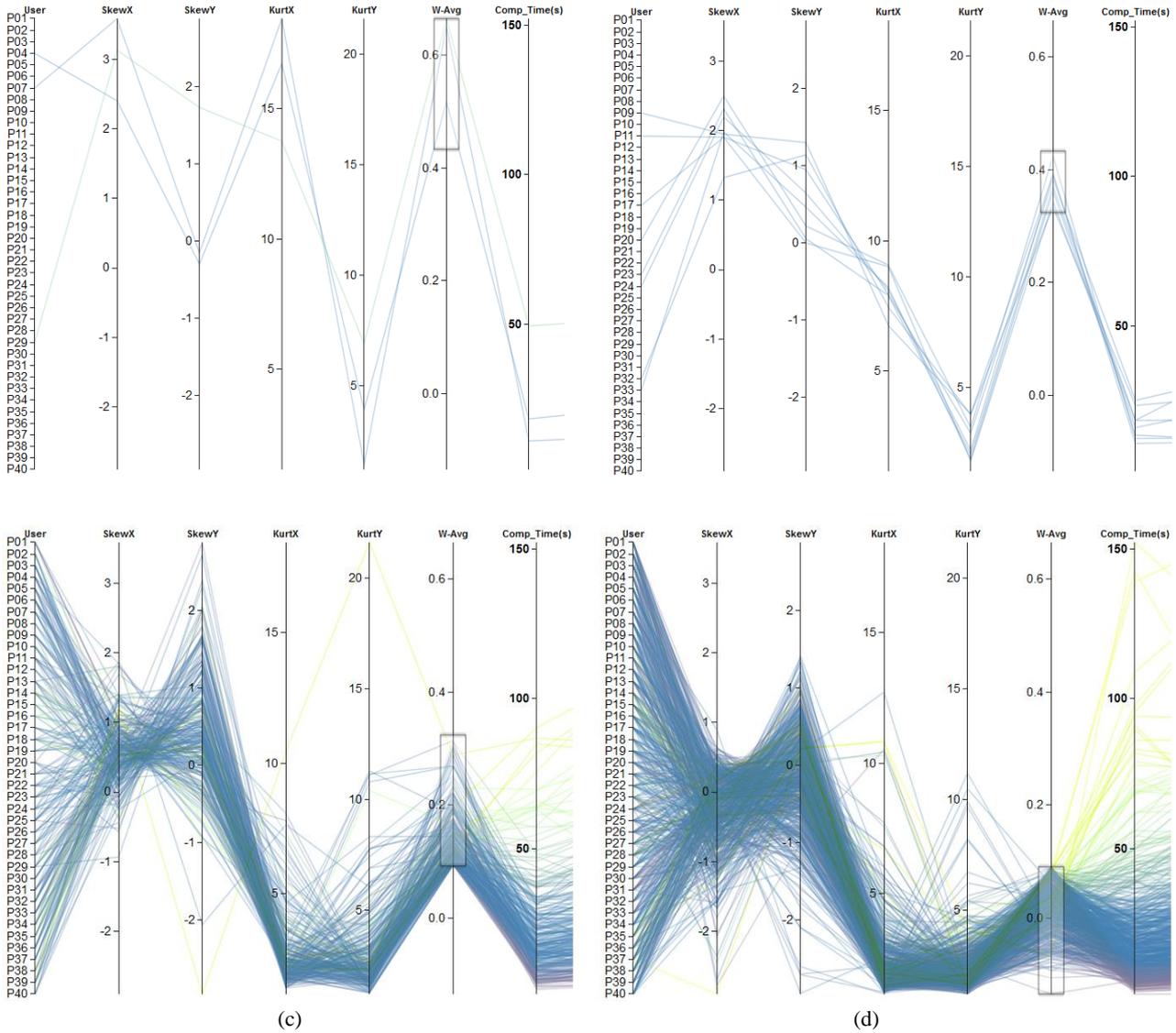

*Figure 12*. Combination of four metrics (*SkewX, SkewY, KurtosisX,* and *KurtosisY*). Color encoding is based on the clustering according to *completion time*. The four images show a rough categorization of the participants into four groups. Metrics values in (a) show large changes between neighboring axes, whereas in (b) there is a visible drop-off in changes. Images (c) and (d) indicate same values with respect to kurtosis but differ in *SkewY*. The weight for skewness was set to 0.3, whereas 0.7 was assigned to kurtosis-related metrics.

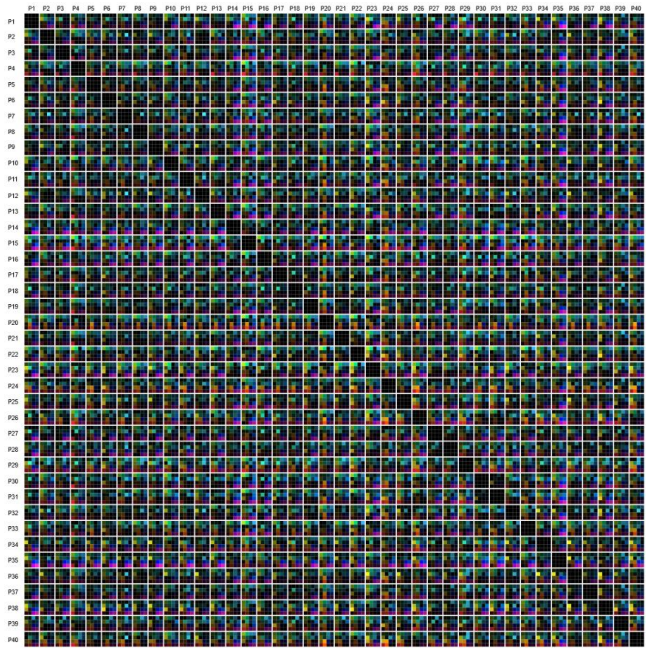

*Figure 13.* Similarity matrices including all participants. All 16 similarity metrics are depicted by color-encoded. The order of the participants is increasing from left to right and top to bottom, according to the participant ID.

During the following *Participant Group Analysis*, our multi-dimensional stacking approach allows us to compare all participants to each other in much detail. Standard approaches generate therefore a similarity matrix where the columns and rows represent the participants, which are order alphanumerically. Each cell of the matrix contains a similarity value. In our case, a matrix cell contains several color-encoded elements, which correspond to the similarity values of the used metrics. An example of this is shown in Figure 13. Here, we can see some isolated group structures, but in general the visualization looks rather chaotic. By performing a reordering of the participants based on the visual inspection in parallel coordinate plots, we are able to introduce, to some degree, order into the chaos: getting similar participants spatially closer reveals more group structures than before.

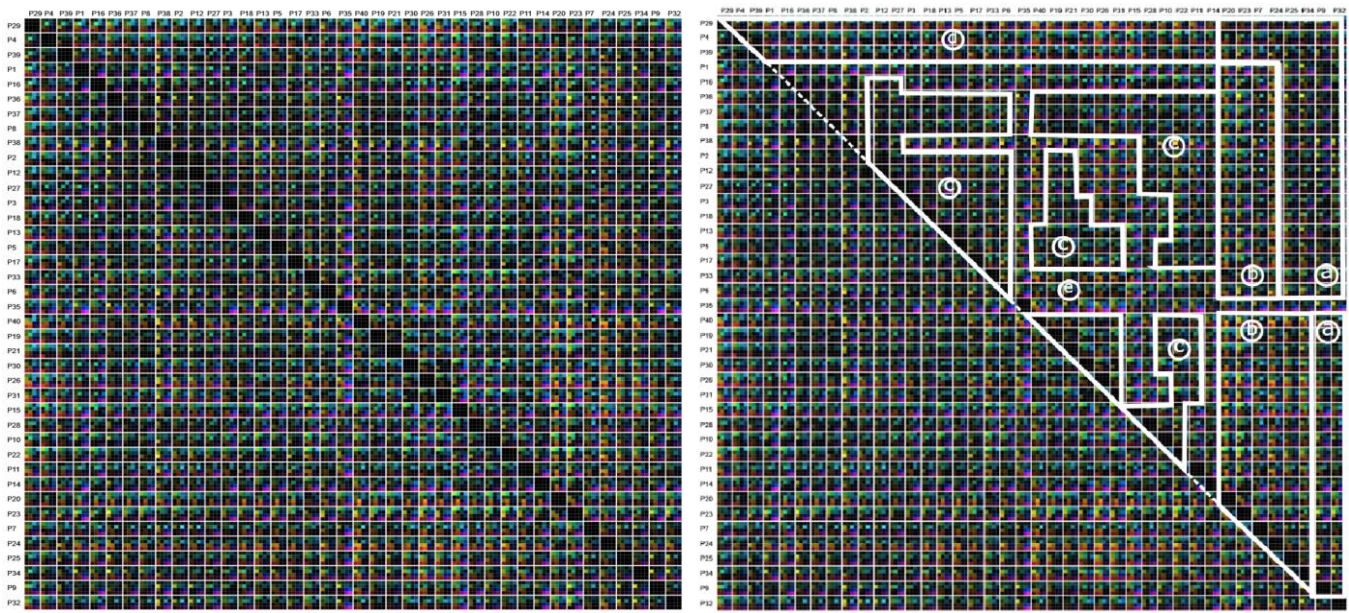

(a)                                                                (b)

*Figure 14.* Similarity matrices including all participants. All 16 similarity metrics are depicted by color. The order of the participants was changed according the result of the clustering based on the new combined metric (skewness in x- and y- direction with a weight of 0.3; kurtosis in x- and y- direction with a weight of 0.7). By visual inspection, six clusters emerge ((a)—(e)), highlighted by white lines.

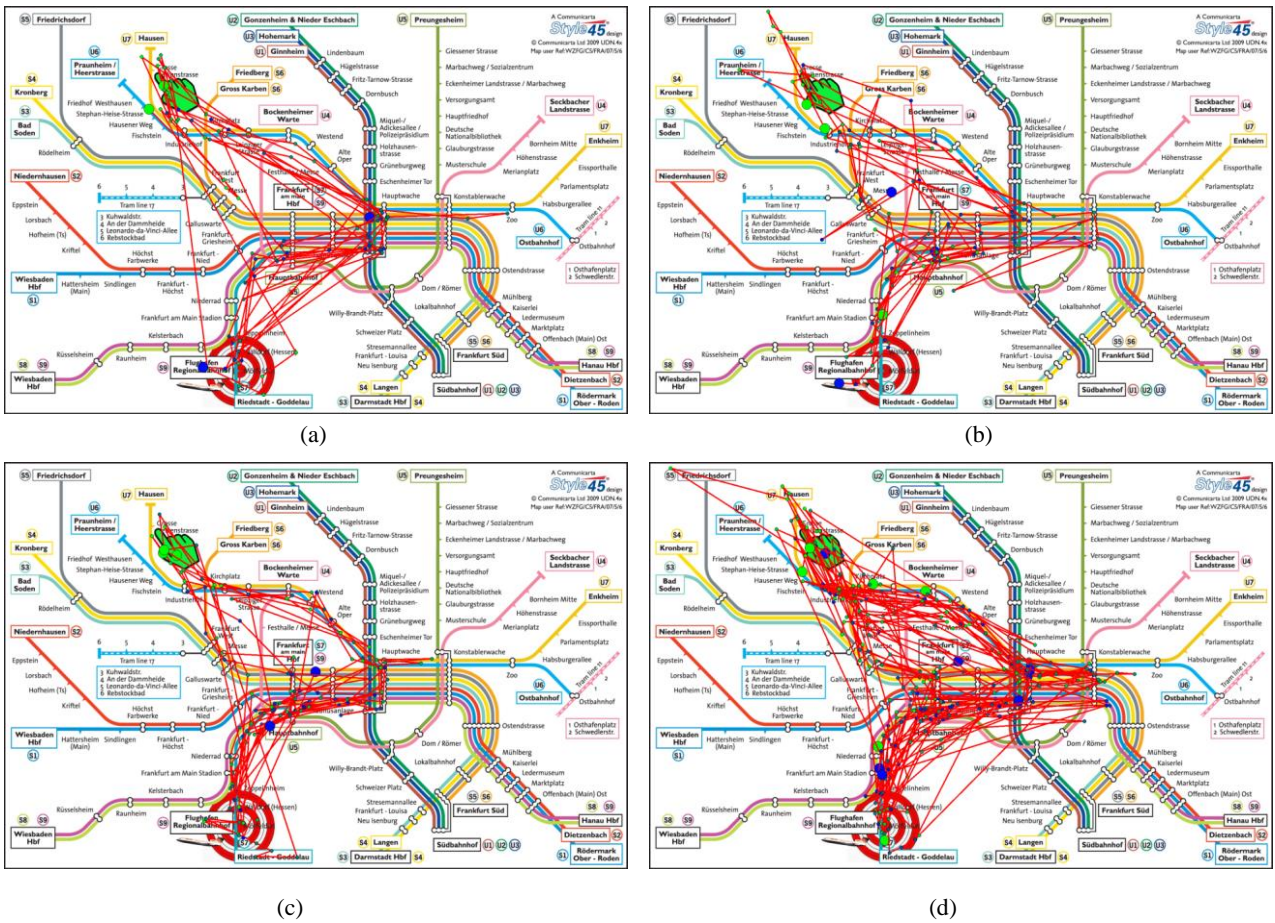

*Figure 15.* Examples of scanpaths for participants that were confronted with color-encoded metro maps. (a) shows scanpaths of participants who are contained in group (b), selected from the horizontal axis. (b) shows scanpaths of participants who are contained in group (a), selected from the horizontal axis. (c) and (d) show scanpaths of participants selected from the vertical axis who are included in group (a) and (b). Scanpaths in (c) are more similar to (a), since they are focused on a path leading from the start to the main station and then to the destination. Scanpaths in (d) are more similar to those in (b), since they are less focused on one specific path involving also areas left and right of the main station. The start location is designated with a green hand and the destination with a target symbol. The start of the scanpath is indicated by a big green dot and the end of the scanpath by a blue dot. The color of smaller fixations in between is gradually changing from green to blue encoding the temporal ordering of the fixations.

Figure 14(a) depicts the similarity matrix with reordered participants, whereas Figure 14(b) highlights a possible grouping. The new order of participants is the result of hierarchical clustering based on a combined metric (skewness in x- and y- direction with a weight of 0.3; kurtosis in x- and y- direction with a weight of 0.7). By using the new ordering, we identified six different groups exhibiting same visually characteristics.

Inspecting the different groups in more detail using scanpaths as part of a qualitative evaluation, we can see that, e.g., group (a) and (b) differ in defining characteristics of scanpaths that were investigated by the participants. Selecting scanpaths of participants from the hori-zontal axis that are within group (b) (see Figure 15 (a)) shows that there is a designated path that was investigated, while scanpaths of participants in group (a) (see Figure 15 (b)) are more diverse. Comparing these scanpaths with those of participants selected from the vertical axis that are part of both groups (see Figure 15 (c) and (b)), we can see that the scanpaths share similar characteristics to the scanpaths of participants chosen from the horizontal axis, thus the scanpaths are separated into two groups exhibiting two defining characteristics: a more focused versus a more diverse or widespread investigation.

## Discussion

We described an approach for analyzing eye-tracking metrics based on recorded eye-tracking data of conducted eye-tracking studies. We provided two visualization techniques - parallel coordinates plots as well as matrix visualization, but we are aware that there is a number of limitations regarding both approaches. In this section, we briefly explain the major problems and challenges that can occur while dealing with this kind of data and visualization techniques focusing on visual and perceptual issues.

If the number of metrics increases, the visualizations can reach their limits. For example, the number of displayed metrics has an impact on the number of axes in parallel coordinates and the vertical stripes between which correlations can be detected. If the number of metrics grows, the gaps become smaller and smaller, leading to visual scalability problems. A growing number of participants leads to many more polylines plotted on top of each other and an increase of visual clutter. A similar effect occurs in the matrix visualization. In case of large number of metrics, the matrix cells should be subdivided into more elements and hence, the display space for each individual one gets smaller. Moreover, if the number of metrics is not a perfect square, then there are some empty elements in the rectangular cells in both rows and columns. If the number of study participants grows, the number of rows and columns in the matrix increases with quadratic growth and, hence, the display space for individual cells decreases rapidly.

Since we have to deal with color coding, we are aware of the fact that too many differentiable groups can lead to perceptual problems, particularly in the case of parallel coordinates, in which many polylines are crossing. A positive aspect of the parallel coordinates is that the vertical axes are easy to use for judging metric values due to their alignment on common scales, as evaluated by Cleveland and McGill (1986). This becomes more problematic in matrix visualizations in which the metric values (or comparisons and relations) are displayed in a color-coded fashion. This means that an analyst has to compare the individual matrix cells and their elements by comparing color hues which is perceptually problematic (Ware, 2012). However, clustering and grouping of the matrix cells helps identify groups of similar behavior reflected in a similar coloring of the cells.

## Conclusion and Future Work

We described a workflow to analyze eye-gaze data consisting of three stages: *Data Preprocessing*, *Metric Analysis*, and *Participant Group Analysis*. We start with preprocessing the raw data into meaningful metrics. The preprocessed data is then visualized with a parallel coordinates plot, which is the basis for metric analysis, reflecting correlation among the metrics. After identifying meaningful metrics, they are used during the *Participant Group Analysis*. Here, we perform clustering based on the selected metrics and their assigned weights to generate groups of participants exhibiting similar behavior, which is the ultimate aim of our paper. The result of clustering is visualized in a dimensional-stacked similarity matrix, where each cell contains multiple elements representing color-encoded similarity values computed for each selected metric. By rearranging the participants based on clustering, we achieve a better perceptual layout for the matrix. Insights gained in this stage can be redirected back to the parallel coordinates plots to refine or change the metrics and their weights to further improve the grouping. We illustrated our visual analytic framework by means of applying it to the data of a formerly conducted eye-tracking study. Finally, we discussed limitations and scalability issues of our work.

For future work, we plan to enable the interaction between both visualization techniques we have used. Changes in metrics and their weights in parallel coordinates will automatically steer the clustering and generation of the dimensional-stacked similarity matrix. To check the usability extent of our tool, we also plan to evaluate our tool by means of an eye-tracking study. Moreover, we plan to experiment with other visualization techniques and further interaction principles with the goal to find further insights in the eye-tracking data and to draw conclusions about the visual scanning strategies of any number of study participants.

## Ethics and Conflict of Interest

The author(s) declare(s) that there is no conflict of interest regarding the publication of this paper.

## Acknowledgements

This research was partially supported by NSF grant IIS 1527200 and the MSIP (Ministry of Science, ICT and Future Planning), Korea, under the "ITCCP Program" (NIPA-2013- H0203-13-1001), directed by NIPA. We also thank the German Research Foundation (DFG) for financial support within project B01 of SFB/Transregio 161. Metro maps in Figure 15 are designed by Robin Woods (www.robinworldwide.com) and licensed from Communicarta Ltd (www.communicarta.com).

## References

Anderson, N. C., Anderson, F., Kingstone, A., & Bischof, W. F. (2015). A comparison of scanpath comparison methods. *Behavior Research Methods*, 47(4), 1377–1392. doi: 10.3758/s13428-014-0550-3

Blascheck, T., Burch, M., Raschke, M., & Weiskopf, D. (2015). Challenges and perspectives in big eye movement data visual analytics. *In Proceedings of Symposium on Big Data Visual Analytics, BDVA* (pp.17–24). doi: 10.1109/BDVA.2015.7314288

Bouguettaya, A., Yu, Q., Liu, X., Zhou, X., & Song, A. (2015). Efficient agglomerative hierarchical clustering. *Expert Systems with Applications,* 42(5), 2785-2797. doi: 10.1016/j.eswa.2014.09.054

Burch, M., Chuang, L., Fisher, B., Schmidt, A., & Weiskopf, D. (2017). *Eye Tracking and Visualization: Foundations, Techniques, and Applications. ETVIS* 2015. Springer. doi: 10.1007/978-3-319-47024-5

Burch, M., Kumar, A., Mueller, K., & Weiskopf, D. *(2016).* Color bands: visualizing dynamic eye movement patterns. *In IEEE Second Workshop on Eye Tracking and Visualization (* pp. 40-44*).* doi: 10.1109/ETVIS.2016.7851164

Chen, H.-C., & Chen, A. L. (2001). A music recommendation system based on music data grouping and user interests. *In Proceedings of the Tenth International Conference on Information and Knowledge Management* (pp. 231–238). doi: 10.1145/502585.502625

Cleveland, W. S., & McGill, R. (1986). An experiment in graphical perception. *International Journal of Man-Machine Studies,* 25(5), 491–501. doi: https://doi.org/10.1016/S0020-7373(86)80019-0

d'Ocagne, M. (1885*). Coordonnées parallèles & axiales: méthode de transformation géométrique et proceed nouveau de calcul graphique déduits de la considération des coordonnées parallèles*. Gauthier-Villars.

Fua, Y.-H.,Ward, M. O., & Rundensteiner, E. A. (1999). Hierarchical parallel coordinates for exploration of large datasets. *In Proceedings of the Conference on Visualization' 99* (pp. 43–50). doi: 10.1109/VISUAL.1999.809866

Graham, M., & Kennedy, J. (2003). Using curves to enhance parallel coordinate visualisations. I*n Seventh International Conference on Information Visualization* (pp. 10–16). doi: 10.1109/IV.2003.1217950

Hauser, H., Ledermann, F., & Doleisch, H. (2002). Angular brushing of extended parallel coordinates. *In IEEE Symposium on Information Visualization* (pp. 127–130). doi: 10.1109/INFVIS.2002.1173157

Heinrich, J., Luo, Y., Kirkpatrick, A. E., Zhang, H., & Weiskopf , D. (2011). Evaluation of a bundling technique for parallel coordinates. *arXiv preprint arXiv:1109.6073.*

Heinrich, J., & Weiskopf, D. (2013). State of the art of parallel coordinates. *In Eurographics 2013 - State of the Art Reports* (pp. 95–116). doi: 10.2312/conf/EG2013/stars/095-116

Holmqvist, K., Nyström, M., Andersson, R., Dewhurst, R., Halszka, J., & Weijer, J. van de. (2011). *Eye tracking: A Comprehensive Guide to Methods and Measure.* Oxford University Press.

Inselberg, A. (1985). The plane with parallel coordinates. *The Visual Computer*, 1(2), 69–91. doi: 10.1007/BF01898350

Krejtz, K., Coltekin, A., Duchowski, A., & Niedzielska, A. (2017). Using coefficient k to distinguish

ambient/focal visual attention during map viewing. *Journal of Eye Movement Research*, 10(2). doi: http://dx.doi.org/10.16910/jemr.10.2.3.

Kumar, A., Netzel, R., Burch, M., Weiskopf, D., & Mueller, K. (2016). Multi-similarity matrices of eye movement data. *In IEEE Second Workshop on Eye Tracking and Visualization* (pp. 26–30). doi: 10.1109/ETVIS.2016.7851161

Kurzhals, K., Burch, M., Pfeiffer, T., & Weiskopf, D. (2015). Eye tracking in computer-based visualization. *Computing in Science and Engineering*, 17(5), 64–71. doi: 10.1109/MCSE.2015.93

Kurzhals, K., Heimerl, F., & Weiskopf, D. (2014). IseeCube: Visual analysis of gaze data for video. *In Proceedings of the Symposium on Eye Tracking Research and Applications* (pp. 43–50). doi: 10.1145/2578153.2578158

LeBlanc, J., Ward, M. O., & Wittels, N. (1990). Exploring n-dimensional databases. *In Proceedings of the 1st Conference on Visualization* (pp. 230–237). doi: 10.1109/VISUAL.1990.146386

Li, X., Çöltekin, A., & Kraak, M.-J. (2010). Visual exploration of eye movement data using the space-timecube. *In International Conference on Geographic Information Science* (pp. 295–309). doi: 10.1007/978-3-642-15300-6_21

Moon, B., Jagadish, H. V., Faloutsos, C., & Saltz, J. H. (2001). Analysis of the clustering properties of the Hilbert space-filling curve. *IEEE Transactions on Knowledge and Data Engineering*, 13(1), 124–141. doi: 10.1109/69.908985

Netzel, R., Ohlhausen, B., Kurzhals, K., Woods, R., Burch, M., & Weiskopf, D. (2017). User performance and reading strategies for metro maps: An eye tracking study. *Spatial Cognition & Computation*, 17(1-2), 39-64. doi: 10.1080/13875868.2016.1226839

Pajer, S., Streit, M., Torsney-Weir, T., Spechtenhauser, F., Möller, T., & Piringer, H. (2017). Weightlifter: Visual weight space exploration for multicriteria decision making. *IEEE Transactions on Visualization and Computer Graphics*, 23(1), 611–620. doi: 10.1109/TVCG.2016.2598589

Paliouras, G., Papatheodorou, C., Karkaletsis, V., & Spyropoulos, C. D. (2002). Discovering user communities on the internet using unsupervised machine learning techniques. *Interacting with Computers*, 14(6), 761–791. doi: /10.1016/S0953-5438(02)00015-2

Raschke, M., Chen, X., & Ertl, T. (2012). Parallel scanpath visualization. *In Proceedings of the 2012 Symposium on Eye Tracking Research and Applications* (pp. 165-168). doi: 10.1145/2168556.2168583

Rosenholtz, R., Li, Y., Mansfield, J., & Jin, Z. (2005). Feature congestion: a measure of display clutter. *In Proceedings of the 2005 Conference on Human Factors in Computing Systems* (pp. 761–770). doi: 10.1145/1054972.1055078

Shi, X., Zhu, J., Cai, R., & Zhang, L. (2009). User grouping behavior in online forums. *In Proceedings of the 15th ACM SIGKDD International Conference on Knowledge Discovery and Data Mining* (pp. 777–786). doi: 10.1145/1557019.1557105

Vernone, A., Berchialla, P., & Pescarmona, G. (2013). Human protein cluster analysis using amino acid frequencies *PloS one*, 8(4), e60220. doi: 10.1371/journal.pone.0060220

Ware, C. (2012). *Information Visualization: Perception for Design*. Elsevier.

Wegman, E. J. (1990). Hyperdimensional data analysis using parallel coordinates. *Journal of the American Statistical Association*, 85(411), 664–675. doi: 10.2307/2290001

West, J. M., Haake, A. R., Rozanski, E. P., & Karn, K. S. (2006). eyepatterns: software for identifying patterns and similarities across fixation sequences. *In Proceedings of the 2006 Symposium on Eye Tracking Research & Applications* (pp. 149–154). doi:10.1145/1117309.1117360

Table 1
Used metrics are divided into three categories. The first group are eye-tracking metrics based on movement measures. The second group are numerosity measures. The third group are statistical (position) measures of spatial distributions of fixation for $x-$ and $y-$coordinates. These statistical measures allow an analyst to get an impression about the most prominent areas within stimuli.

| Eye-Tracking Metric | Definition |
| --- | --- |
| Average fixation duration | It is often used as an indicator for the cognitive processing depth. High values typically mean that a participant spent more time thinking about an area, for example, due to high complexity of the scene or an absence of intuitiveness in it. Low values in a local area can be the result of stress. |
| The average saccade length | Also called saccade amplitude. A long saccade length can be interpreted as an explorative eye movement, whereas short saccade lengths may occur when the task difficulty increases as short eye movements are used to collect information from a restricted area to support the current cognitive process. |
| Average saccade duration | Average time to move from one fixation to another and therefore the average time with no visual intake. Average saccade duration is decreasing for more difficult tasks, as well as with a decreased processing capacity. |
| $\mathcal{K}$ coefficient | Characterizes dynamics of ambient and focal attention per individual scanpath (Krejtz, Coltekin, Duchowski, & Niedzielska, 2017). Positive values indicate longer fixation durations and shorter saccades: focal attention. Negative values indicate the opposite: longer saccades and shorter fixations, therefore ambient attention. In the ambiguous case, where the coefficient is zero, subjects could have made either long saccades followed by long fixations or short saccades followed by short fixations. Therefore, distinguishing ambient and focal attention is not possible in this case |
| Number of fixations | General measure that can be applied to specifically defined areas or the whole stimulus, and it is correlated to the time spent in an associated area. The combination of spent time and number of fixations could be used to find different kinds of behavior, e.g., areas that exhibit the same number of fixations, but different spent time indicate a different behavior, possibly influenced by the content of the associated area. |
| Fixation rate | Is roughly proportional to the inverse of the average fixation duration. This metric can be used to interpret task difficulty. Furthermore, it is used to predict target density and therefore an indicator for measuring mental workload. |
| Number of saccades | Proportional to the number of fixations and related to the fixation duration. If the number of saccades is increasing, in a fixed amount of time, the fixation duration is decreasing. |
| Saccade rate | In general, almost identical to the fixation rate. It is an indicator for mental workload, arousal, and fatigue. The saccade rate is decreasing with an increase of task difficulty, mental workload, or fatigue. An arousal leads to an increase of the saccade rate. |
| Scanpath length | It is the sum of the length of all saccades. |
| Completion time | Time measured between the start of displaying a stimulus and the participant finishing the task. |
| Standard deviation ($x$ & $y$) | Used to quantify the amount of variation or dispersion of a distribution. Lower values indicate that the values of the distribution are closer to the mean value, whereas larger values of the distribution indicate a higher dispersion. |
| Skewness ($x$ & $y$) | Measures the asymmetry of a distribution, i.e., whether the left tail of a unimodal distribution (negative skewness) or the right tail (positive skewness) is longer or fatter. |
| Kurtosis ($x$ & $y$) | Like skewness, it describes the shape of a distribution. It is the moment of the distribution and tells an analyst the reasons for the variance within the data. Higher kurtosis indicates infrequent extreme deviations, whereas lower kurtosis indicates frequent modestly sized deviations. |